\documentclass[12pt,a4paper]{article}

\usepackage[english]{babel}

\usepackage{epsfig}
\usepackage{dcolumn}
\usepackage{amsmath}
\usepackage{cite}
\usepackage{changebar}

\makeatletter \@addtoreset{equation}{section}
\makeatother 
\renewcommand{\thefootnote}{\alph{footnote}}

\newcommand{\Tr} {\mbox{Tr}}

\begin{document}
\thispagestyle{empty}
\hbox{}
%
 \mbox{} \hfill IRB-TH 4/04\hfill

 \mbox{} \hfill BI-TP 2004/20\hfill

\begin{center}
\vspace*{1.0cm}
\renewcommand{\thefootnote}{\fnsymbol{footnote}}
{\huge \bf  ENTROPY FOR DIQUARKS IN EXOTIC QUARK STATES}\\

\vspace*{1.0cm}
{\large David E. Miller$^{1,2,3}$}
\\
\vspace*{1.0cm}
${}^1$ Division of Theoretical Physics, Rudjer Bo$\check{s}$kovi\'c Institute,
\\ HR-10002 Zagreb, Croatia\\
${}^2$Fakult\"at f\"ur Physik, Universit\"at Bielefeld, Postfach 100131,\\
D-33501 Bielefeld, Germany 
\footnote{email: dmiller@physik.uni-bielefeld.de} \\
${}^3$ Department of Physics, Pennsylvania State University,
Hazleton Campus,\\
Hazleton, Pennsylvania 18201, USA 
\footnote{permanent address, email: om0@psu.edu} \\
\vspace*{2cm}
{\large \bf Abstract}

We discuss the quantum state structure on the basis of $SU(3)_c$
for some known exotic quark systems using a model which describes 
these particular states as  highly correlated diquarks and 
antidiquarks. We are then able to calculate for a single colored 
diquark a finite von Neumann entropy from the quantum reduced 
density matrix, from which we explicitly evaluate the likelihood 
of certain arrangements of the quark flavors in a given diquark. 
These results can be related to some recently experimentally found 
pentaquark systems as well as a possible model for the scalar mesons. 
\end{center}

\vfill
\newpage


\section{\bf Introduction}

\noindent
The recent experimental discovery of the pentaquark states~\cite{pent1,pent2}
has brought about new intensity to the search for other exotic quark states
beyond the usual meson and baryon singlet quark structures. As well 
as this experimental work there has been recent numerical work on 
particular cases involving the diquark structure~\cite{NakaSait} which
provides some information on the nature of the interaction between the 
quarks within a diquark. In the recent past it has proven itself 
quite useful to have such information for the case of the 
scalar meson~\cite{AlfJaf} since the states are experimentally 
rather difficult to determine their exact physical properties. 
Furthermore, much earlier there had been important theoretical 
work based on the bag model which has provided a basis for the study 
of various exotic quark structures. The work of Jaffe~\cite{Jaf} 
used the bag model to study the dimeson state given in the form 
$q^2{\bar{q}}^2$ as a diquark and an antidiquark combination. 
Further work on the exotic quark formation by Strottman~\cite{Strot} 
has concentrated particularly on the states $q^4{\bar{q}}$ and 
$q^5{\bar{q}}^2$, which are respectively the pentaquark 
and the baryon-diquark-antidiquark combined state. \\
~\\
\indent
     In the present work we want to discuss the role of the colored 
diquarks~\cite{diquark} as a basic element of the exotic systems. 
Furthermore, we shall suggest why the diquark is very basic to the 
structure of baryons, whereby we maintain that we may consider that 
baryon is a diquark and a quark. An important recent work
by Jaffe and Wilczek~\cite{JafWil} has shown a significant relation 
of the diquark structure to the pentaquark states. Because of this
proposed model we shall treat the diquark as an important structure 
in the determination of the exotic states. As we did in our earlier 
treatment, we shall especially look at the role of the entropy~\cite{vNeu} 
relating to the color structure of the quantum ground state. Quite 
recently we have looked at the quantum entropy $S_q$ of the color
singlet quarks within the hadrons~\cite{Mil} by calculating that 
of the quarks in the singlet meson and baryon states. Afterwards, 
this calculation was used to calculate the effect of the contributions 
of $S_q$ on the equation of state for baryons with the same quark flavors
in a bag model situation~\cite{MilTaw}.\\
~\\
\indent
     The  standard model has the color charge carried by the quarks as
the fundamental property of the strong nuclear interaction~\cite{DoGoHo}. 
In clear contrast to the other known charges the color charge cannot be 
easily isolated and separately measured. In nature it always appears as
part of selective states of the $SU(3)_c$, wherein the quarks and antiquarks
are placed respectively in the fundamental $~\bf{3}~$ and antifundamental 
$~\bf{\bar3}~$ representations of this group. These two representations 
together with the adjoint representation of $SU(3)_c$ make up the 
symmetry structure of Quantum Chromodynamics(QCD)~\cite{Mut,Kug}. The two 
main categories of strong interacting particles (hadrons) are the mesons, 
which may be written as a product of the fundamental and the antifundamental 
representations $~\bf{3}~{\otimes}~\bf{\bar{3}}~$ and the baryons, 
which are a product of three fundamental representations 
$~\bf{3}~{\otimes}~\bf{3}~{\otimes}~\bf{3}~$.  Although the different quarks 
have other properties like spin, electrical charge, mass as well as a very
special property called {\it{flavor}}, which we shall see has a very
important role in the determination of the size of the entropy 
in relation to the diquark structure.\\
~\\
\indent 
     In this work we shall write the quark and antiquark color states as 
follows: $~|0\rangle~$,$~|1\rangle~$,$~|2\rangle~$ and 
$~|\bar{0}\rangle~$,$~|\bar{1}\rangle~$,$~|\bar{2}\rangle~$. We shall 
use this notation to describe the orthonormal bases of the fundamental 
and the antifundamental representations of $SU(3)_c$ instead of the 
more common color names. From these color basis states we can construct
a representation for the color exotic state functions-- in particular for
the dimeson ${\Psi}_{dm}$ and pentaquark ${\Psi}_{pe}$ groundstates.
From these color state functions we are able to construct the corresponding
density matrices ${\bf{\rho}}_{dm}$ and ${\bf{\rho}}_{pe}$ for the color
states following the usual prescription given by quantum mechanics~\cite{LaLi}. 
The general procedure for the following work is that we shall arrive at the
single diquark reduced density matrix ${\bf{\rho}}_{dq}$, which is of
particular interest in all further calculations. From ${\bf{\rho}}_{dq}$
we can directly calculate the quantum entropy in the sense of von Neumann
~\cite{vNeu,LaLi}. As in our previous calculations~\cite{Mil,MilTaw}, 
the results of this calculation show a significant
contribution of order one to the entropy of the quarks in the exotic
states. This value is given as a pure number without 
physical dimensions when we use the usual high energy units with 
$\hbar$, $c$ and Boltzmann's constant $k$ all set to the value one.\\
~\\
\indent 
     The further implications of these results can be brought together 
with some of our earlier work on the hadronic states~\cite{Mil,MilTaw}.
We start by combining the known singlet hadronic states. Since the  
lowest scalar meson, which was formerly called the $\sigma$-particle,
is known to decay into two pions at very low energies, but also can 
include two kaons at energies\footnote[1]{The actual scalar meson 
states are discussed in the "Review of Particle Physics"~\cite{PDG} in
a "Note on Scalar Mesons" pp. 506 -510 and 522 -526. The actual scalar 
states go under the names $f_0(600)$, $f_0(980)$ and $a_0(980)$. These
states are all known to decay into pairs of pseudoscalar mesons and, of
course, pairs of photons.} just below $1 GeV$. Then the scalar meson 
can be constructed from the two $q{\bar{q}}$ combinations as the direct 
product state which gives the obvious final state of two pseudoscalar 
mesons with the necessary positive parity. From this consideration a 
dimeson is just a simple reordering of the quark and antiquark color 
states which clearly uphold this singlet structure leading to the colorless 
state function ${\Psi}_{dm}$ with  the quark structure of $q^2{\bar{q}}^2$. 
Similarly, the pentaquark states are built from the color states of the 
decay into the baryon and meson final states. Here the $q^4\bar{q}$ are 
structured as two diquarks and an antiquark. The five color states only 
maintain their colorlessness with the total baryon color wave function. 
Then ${\Psi}_{pe}$ is colorless even when each single contributiion is 
colored. In the next section we shall write out these state function 
explicitly and indicate the forms of the respective density matrices. 
Afterwards we can find the value of the entropy in the different cases. 
Finally we conclude with a few remarks on the exotic structures.

\section{\bf Quark Structure of the Exotic States}

\noindent
The starting point for the investigation of the ground state 
of the quark structure of the exotic states is the evaluation 
of the density matrix~\cite{LaLi} for the singlet state quark 
structure of the mesons and baryons~\cite{Mil,MilTaw}. This known 
color structure may be put together in specific combinations to 
yield the known final states from the exotic states' decays.\\ 
~\\
\indent
     Here we quickly recall how the singlet meson and 
baryon color state functions~\cite{DoGoHo} are constructed. 
The meson is simply the sum of the direct product of quark
and antiquark states of the form 
$\frac{1}{\sqrt 3}\sum_i|i\bar i\rangle$, where the numerical 
forefactor serves as the correct normalization. Similarly the 
baryons color states are written as a direct product of three 
colored quark states where the permutation $P\{ijk\}$ determines 
the sign in front of each term in the sum. The baryon color singlet 
state is given by $\frac{1}{\sqrt 6}\sum_{P}(-1)^P|ijk\rangle$.
The resulting entropy for both the meson and baryon colored states 
has already been calculated~\cite{Mil} to yield $S_q = \ln 3$.\\
~\\
\indent
     Here we shall consider explicitly only the color parts 
of the dimesons' state functions ${\Psi}_{dm}$ and its 
conjugate ${\Psi}^{*}_{dm}$. The dimesons are constructed from 
the direct product of the fundamental representations of the 
constituents in the form  
$~\bf{3}~{\otimes}~\bf{3}~\otimes~\bf{\bar{3}}~\otimes~\bf{\bar{3}}$. 
Thus the color singlet state functions for the dimesons are taken from
the states arising out of this product representation. We are not able
to represent the dynamical process which takes place in the decay to
two or more meson singlet states. We notice that each entry is itself
color neutral so that in combination these states make up the nine
basic contributions to the color singlet dimeson states. Thus we can
expect that these are pure physical states even if they may not be
easily separated and measured. We write these in the following form:   
\begin{eqnarray}
 {\Psi}_{dm}&=&{\frac{1}{{3}}}(|0~0~\bar{0}~\bar{0}\rangle~+~
               |0~1~\bar{0}~\bar{1}\rangle~+~|1~0~\bar{1}~\bar{0}\rangle~+~
               |0~2~\bar{0}~\bar{2}\rangle~+~|2~0~\bar{2}~\bar{0}\rangle~+~
               \nonumber \\
            & &|1~1~\bar{1}~\bar{1}\rangle~+~|1~2~\bar{1}~\bar{2}\rangle~+~
               |2~1~\bar{2}~\bar{1}\rangle~+~|2~2~\bar{2}~\bar{2}\rangle).
  \label{dimesonstate}
\end{eqnarray}
Here we have kept the left to right ordering of the quarks and antiquarks.
Also for the corresponding conjugate singlet state function of the dimeson 
we maintain a similar form\footnote[2]{The order of the representation remains
the same although the "ket" states $|ij..\rangle$ are replaced by the "bra" 
states $\langle ij..|$ so that again "i", "j".. are the first, second.. 
particle.} that contains the corresponding conjugate color state vectors 
as previously discussed for the simple meson~\cite{Mil}. Thus we need 
not explicitly rewrite these terms for the conjugate state here.\\
~\\
\indent
    Similarly we may write the state function for pentaquarks ${\Psi}_{pe}$ 
coming from the singlet baryon and meson structures in the representation 
$~\bf{3}~{\otimes}~\bf{3}~{\otimes}~\bf{3}~{\otimes}~\bf{3}~{\otimes}~\bf{\bar{3}}$.
Again the same ordering principle is kept for the quark and antiquarks.
However, for the pentaquark each state is not colorless but only within
the cycles of the permutation of the first three colors, which means
that there are three subgroupings of colorless states. Thereby we may
write down the four quarks and one antiquark terms of the 
pentaquarks, which can be set out as follows:
\begin{eqnarray}
 {\Psi}_{pe}&=&{\frac{1}{\sqrt{18}}}(|0~1~2~0~\bar{0}\rangle~+~
               |1~2~0~0~\bar{0}\rangle~+~|2~0~1~0~\bar{0}\rangle~-~
               |0~2~1~0~\bar{0}\rangle~-~|1~0~2~0~\bar{0}\rangle~-~
               \nonumber\\
            & &|2~1~0~0~\bar{0}\rangle~+~...~-~|2~1~0~2~\bar{2}\rangle),
  \label{pentsing}
\end{eqnarray}
where the dots indicate the other eleven missing terms arising from the 
further mesonic color combinations. Here we have written only the six
states in the first colorless subgrouping. In an analogous way to 
that explained for the dimeson we may also express the corresponding
conjugate state function in the order of the tensor product 
for the pentaquarks using the correct conjugate color states.\\

\section{\bf Density Matrix Reduction for the Exotic States}

\noindent
We can now write down the density matrices $\bf{\rho}$ for the exotic states 
using the direct product of ${\Psi}$ and ${\Psi}^{*}$ of the color state
functions. This gives for the color singlet dimesons ${\bf{\rho}}_{dm}$
and pentaquark ${\bf{\rho}}_{pe}$  the density matrices in the 
following general forms:
\begin{equation}
{\bf{\rho}}_{dm}~=~{\Psi}_{dm} {\Psi}^{*}_{dm}
  \label{eq:denmeson}
\end{equation}
and
\begin{equation}
{\bf{\rho}}_{pe}~=~{\Psi}_{pe} {\Psi}^{*}_{pe} .  
  \label{eq:denbaryon}
\end{equation}
~\\
\indent
     Previously we had only considered the ordinary hadronic states as being
made out of the simple singlet combinations of the quark and antiquark 
states~\cite{Mil,MilTaw}. Now we shall extend this study in order to
construct more general combinations of the quarks and antiquarks in
the formulation of the resulting density matrices. We know that 
for the hadrons the wavefunctions yield $\it{pure}$ states~\cite{LaLi}.
Since the exotic states are built up from a direct product of the pure 
hadronic states into which the decay, the resulting density matrices
also again represent pure states. However, the diquarks are a 
constituent subsystem making up the complete exotic state. For the
calculation of the internal quark structure we look at the 
diquark $\it{reduced}$ density matrices, which give the actual
statistical state of the individual diquark within the total system. 
In order to get the reduced density matrices for the diquark in the 
dimesons, we project out all the antidiquark states, where colors 
are specified by $i$ and $j$ in the diquark color pair $[ij]$,
$~\langle [\bar{ij}]|$and $~|[\bar{ij}]\rangle~$ by using the
orthonormality and the completeness properties. Similarly for the 
pentaquark structures we project out first the antiquark state
$~\langle \bar{i}|$ and $~|\bar{i}\rangle$. Then in the second
reduction of the density matrices we project $~\langle [ij]|$ 
and $~|[ij]\rangle$ onto the other two diquark states to the right.  
These diquark states result in two contributions for each color. 
After carrying out these reductions, we find that both the 
reduced density matrices from the dimeson and the pentaquark
for the diquark states take on the same general diagonal form:
\begin{equation}
\label{eq:reddenmes}
{\bf{\rho}}_{dq}~=~{\frac{1}{N}}\sum_{[ij]}\{|[ij]\rangle \langle [ij]|\}.
\end{equation}
\noindent
This is the reduced density matrix for the diquarks in the colored states.
It always yields a completely mixed state with the same eigenvalue 
$\lambda_i = 1/N~$ for each diquark contribution.
~\\
\indent
     In order to determine the exact value of $~N~$ in $~\bf{\rho}_{dq}~$,
we must specify the diquark more exactly. It depends upon the $\it{flavors}$
in the diquark as well as from where it came. In all cases if the two 
quarks in the diquark have the same flavor, the the value of $~N=3~$. 
Two quarks with same color and flavor act as identical particles for which
the Pauli exclusion principle automatically applies up to the spin.
However, with different flavors in the diquark the two quarks are
clearly distinguishable. This is essential to the diquark states in the
density matrix. When the flavors are different the state $|01\rangle$ 
is different from the state $|10\rangle$. Then the projection in the
density matrix, $\langle{10|01}\rangle$ is zero for different flavors, 
but is one for the same flavors. This fact eliminates the nondiagonal
projections for diquarks with different flavors within thereby changing
the eigenvalues. After we have carried out the reduction of the density
matrix for the same flavors\footnote[3]{Here the diquark reduction of the
exotic state is analogous to the first reduction in the baryon with
the same flavors. Then the first reduction leaves the same diquark structure 
which is then reduced to the diagonal singlet density matrix all with the
probability 3.}, we have found that the value $~N=3~$ is then the
same as in the case of the single quark reduced density matrix $~\rho_q~$
for quarks in the hadronic singlet state~\cite{Mil}. However, for the 
dimeson with different flavors the value of the normalization $~N=9~$ 
since there are nine different diquark states which contribute. 
Nevertheless, for the pentaquark state the baryonic states determine 
the nature of the diquarks. Thus we find after the second reduction 
that the value for $~N=6~$ when there are different flavors. In some
ways this result for the pentaquark is surprising since it has many
more states initially than the dimeson. In the pentaquark case the 
reduction of the density matrix depends on the baryon, which with
different flavors in the diquarks also yields $~N=6~$. However, the
dimeson has the diagonal structure of the meson. When there are 
different flavors for the diquarks in the reduced density there are
nine different states.\\

\section{\bf Entropy for Exotic Quark States} 

We can calculate the entropy $~S~$ of the quantum states using the
prescription of von Neumann~\cite{vNeu,LaLi}, which makes direct
use of the density matrix $~\bf{\rho}$. It is simply written as
\begin{equation}
\label{eq:entropy}
                   S~=~-{\Tr(~\bf{\rho}}~{\ln{\bf{\rho}})},
\end{equation}
where the trace "Tr" is taken over the quantum states. When, as is
presently the case, the eigenvectors are known for $~\rho~$, we may
write this form of the entropy in terms of the sum of the 
eigenvalues $~{\lambda}_i$ as follows: 
\begin{equation}
\label{eq:enteig}
                   S~=~-\sum_{i}{{\lambda}_i{\ln{\lambda}_i}}.
\end{equation}
It is obviously important to have positive eigenvalues. For a zero
eigenvalue we use the fact that $x{\ln x}$ vanishes in the small $x$ limit.
Then for the density matrix $~\rho~$ we may interpret ${\lambda}_i$ as the
probablitiy of the state $~i~$ or $~p_i~$. This meaning demands that
$~0~<~p_i~\leq~1~$. Thus the orthonormality condition for the given states
results in the trace condition
\begin{equation}
\label{eq:tracecond}
                  \Tr~{\rho}~=~\sum_{i} p_i~=~1.
\end{equation}
This is a very important condition for the entropy.
~\\
\indent
     We now apply these definitions to the entropy for the quark states.
It is clear that the original hadron states are pure colorless states
which possess$~\it{zero}$ entropy. For the dimeson it is immediately obvious 
since each colored diquark state has the opposing colored antidiquark state 
for the resulting colorless singlet state. The sum of all the cycles determine
the colorlessness of the pentaquark singlet state thereby also yielding no 
entropy.
However, the reduced density matrix for the individual quarks (antiquarks)
$~{\rho}_q~$ or $~{\rho}_{\bar{q}}~$ has a finite entropy. For $SU(3)_c$
all the eigenvalues $~{\lambda}_i~$ in Equation (3.2) have the same value $1/N$.
Thus we find for all the quarks (antiquarks) in these exotic  singlet states
\begin{equation}
\label{eq:entquark}
                  S_{dq}~=~\ln N.
\end{equation}
\indent
     Our further consideration involves the above mentioned different cases. 
When both the quarks in the diquark have the same flavor, then, as stated
above, the diquark reduces to the single quark with $~N=3~$. It follows that
$~S_{dq}= \ln 3$ is the same as $S_q~$ Thus a diquark
in the singlet color state with no specified spin orientations for its 
constituent quarks only can exist with mutually different flavors. In the
pentaquarks with different flavors in the diquarks we have found that $~N=6~$.
This is actually the same result we get for the the usual baryon~\cite{Mil} 
after carrying out the first reduction of the density matrix with different
quark flavors, which yields for $S_{dq}$ just $~\ln 6$. The dimeson with each 
diquark and antidiquark containing internally different flavors has $~N=9~$. 
Thereupon, the color entropy $S_{dq}$ of a diquark with different flavors within 
has the same value for the diquark entropy as the meson with the total quark 
and antiquark entropy of just $~\ln 9$.\\
~\\
\indent
     Hereupon, we may discuss the entropy in some more detail for the main examples 
of the colorless exotic ground states-- the dimesons and the pentaquarks. 
As we have discussed above for the density matrix, all the dimesons 
consist of a diquark-antidiquark pair bound together as a sum of
all the three colors. Since each single quark or antiquark state is equally
weighted in the reduced density matrix, therefore each state posesses equal 
probability of 1/3. Thus we easily get the entropy of $~\ln~3~$. The baryon has 
the doubly reduced density matrix for each single quark state appearing twice so 
that with the normalization factor of 1/6 the probability of each colored quark 
state is again 1/3, which yields the same result for the entropy, $~\ln~3~$. 
This value gives the maximal entropy for a completely  mixed state. For the
pentaquark with different flavors we can carry out the two reductions-- first
the antiquark reduction and then the second diquark reduction. After these
reductions there are exactly six diquark states which all remain in a diagonal
form.

\section{\bf Conclusions}

\noindent
Our objective in this work was to investigate the diquark 
structure of the color states for the quark and antiquark 
systems involving exotic quark states. We chose to discuss two 
obvious examples of these exotic states both from their simplicity 
and their physical relevance. Although there has been recently
a considerable amount of both experimental and theoretical work
on both these cases, it has generally investigated the flavor,
charge or spin structure, but to our knowledge not explicitly 
the color states. It is clear that the experiments do not
have the direct access to the color states which is the case
for the flavor and spin. However, the symmetry rules and the
probability factors arise out of the color states. Thus we have 
seen that the diquark structure depends upon the relation of the
colors to the flavors present. It is the entropy $S_{dq}~=~\ln N$ 
which determines the relation of particular diquark configurations
to the number of nontrivial eigenstates of the reduced density matrix.
This fact allows us to determine the more favorable diquark combinations
for any given exotic quark state.\\
~\\
\indent
     Here we have used a known method in quantum statistical physics 
to examine a situation in high energy particle physics. It can be well 
carried out for the ground state of known particle quark structures in 
the color basis states. Thus the dependence on type of the quark color 
structure in relation to the different flavors is the extension 
of this approach of the three basis states of the fundamental 
representations of $SU(3)_c$. The results of our calculation are
consistent with the expected diquark structure which consist of
colored quarks with different flavors~\cite{JafWil}.\\
~\\

{\bf\Large Acknowledgements}

The author would like to thank Tome Anti$\check{c}$i\'c, Ivan Dadi\'c
and Kre$\check{s}$o Kadjia for many\\ 
essential discussions.
He is also very grateful to the Pennsylvania State University 
Hazleton for the sabbatical leave of absence
and to the Fakult\"at f\"ur Physik  der Universit\"at Bielefeld. 
He thanks the Fulbright Scholarship for support at the Rudjer 
Bo$\check{s}$kovi\'c Institute in Zagreb, Croatia

\end{document}